\documentclass[twocolumn]{aastex631}

\usepackage{amsmath}
\usepackage{multirow}
\usepackage[normalem]{ulem}
\usepackage{hyperref}

\newcommand{\sub}[1]{_{\mathrm{#1}}}
\newcommand{\msun}{M\sub{\sun}}
\newcommand{\nbody}{$N$-body }
\newcommand{\sersic}{S{\'e}rsic }

\def\equationautorefname~#1\null{Eq.~(#1)\null}
\def\figureautorefname~#1\null{Fig.~#1\null}
\newcommand{\appref}[1]{\hyperref[#1]{Appendix~\ref{#1}}}

\begin{document}

\title{Testing the galaxy collision induced formation scenario for the trail of dark matter deficient galaxies with the susceptibility of globular clusters to the tidal force}

\correspondingauthor{Go Ogiya}
\email{gogiya@zju.edu.cn}

\author[0000-0002-3496-8592]{Go Ogiya}
\affiliation{Institute for Astronomy, School of Physics, Zhejiang University, Hangzhou 310027, China}
\author[0000-0003-3236-2068]{Frank C. van den Bosch}
\affiliation{Department of Astronomy, Yale University, PO. Box 208101, New Haven, CT 06520-8101}
\author[0000-0001-6879-9822]{Andreas Burkert}
\affiliation{Universit\"ats-Sternwarte M\"unchen, Scheinerstra\ss e 1, D-81679 M\"unchen, Germany}
\affiliation{Max-Planck-Institut f\"ur extraterrestrische Physik, Postfach 1312, Gie\ss enbachstra\ss e, D-85741 Garching, Germany}
\author[0000-1111-1111-1111]{Xi Kang}
\affiliation{Institute for Astronomy, School of Physics, Zhejiang University, Hangzhou 310027, China}

\begin{abstract}
  It has been suggested that a trail of diffuse galaxies, including two dark matter deficient galaxies (DMDGs), in the vicinity of NGC~1052 formed because of a high-speed collision between two gas-rich dwarf galaxies, one bound to NGC~1052 and the other one on an unbound orbit. The collision compresses the gas reservoirs of the colliding galaxies, which in turn triggers a burst of star formation. In contrast, the dark matter and pre-existing stars in the progenitor galaxies pass through it. Since the high pressures in the compressed gas are conducive to the formation of massive globular clusters (GCs), this scenario can explain the formation of DMDGs with large populations of massive GCs, consistent with the observations of NGC~1052-DF2 (DF2) and NGC~1052-DF4. A potential difficulty with this `mini bullet cluster' scenario is that the observed spatial distributions of GCs in DMDGs are extended. GCs experience dynamical friction causing their orbits to decay with time. Consequently, their distribution at formation should have been even more extended than that observed at present. Using a semi-analytic model, we show that the observed positions and velocities of the GCs in DF2 imply that they must have formed at a radial distance of 5-10\,kpc from the center of DF2. However, as we demonstrate, the scenario is difficult to reconcile with the fact that the strong tidal forces from NGC~1052 strip the extendedly distributed GCs from DF2, requiring 33-59 massive GCs to form at the collision to explain observations.
\end{abstract}


%%%%%%%%%%%%%%%%%%%%%%%%%%%%%%%%%%%%%%%%%%%%%%%%%%%
\section{Introduction} 
\label{sec:intro}

Since the striking report by \cite{vanDokkum2018_df2} that an ultra diffuse galaxy in the group of a large elliptical galaxy, NGC~1052, NGC~1052-DF2 (hereafter DF2) lacks dark matter (DM) mass by a factor of several hundred compared to the expectation by the standard model of galaxy formation and evolution, DF2 has been intensively investigated. Subsequently, a second dark matter deficient galaxy (DMDG), NGC~1052-DF4 (hereafter DF4), was discovered in the same galaxy group \citep{vanDokkum2019}. The proposed formation scenarios for the DMDG population include violent tidal stripping by the host galaxy \citep[e.g.][]{Ogiya2018,Maccio2021,Moreno2022} and galaxy formation in the tidal debris or arms formed in galaxy interactions \citep[][and references therein]{Bournaud2007,Lelli2015,Fensch2019_tdg}. 

\cite{vanDokkum2022_trail} recently discovered a trail of galaxies with low surface brightness, including DF2 and DF4, in the vicinity of NGC~1052, while the membership of individual diffuse galaxies to the NGC~1052 group has not been confirmed yet. To explain the galaxy trail, they proposed the following scenario: About 8\,Gyr ago, an interloper galaxy that had abundant gases collided with a gas-rich satellite galaxy of NGC~1052 with the relative velocity of $\sim 350$\,km/s, comparable to the maximum circular velocity of NGC~1052. The high-speed galaxy collision strongly compresses the gas reservoir of the colliding galaxies and induces a burst of star formation. Meanwhile, collisionless components of the galaxies, i.e., DM and pre-existing stars, pass through the gas. Thus the galaxy collision separates the gas and newly born stars from the DM, forming DMDGs \citep{Silk2019, Lee2021, Trujillo-Gomez2021}. The compressed gas is subsequently stretched out by the tidal force of the passing DM to form a cylindrical structure. After radiative cooling, the Jeans instability drives the fragmentation of the filament to form multiple DMDGs on a line \citep{Shin2020,vanDokkum2022_trail}. As the considered process resembles the formation of the bullet cluster \citep{Clowe2006}, we refer to this scenario as the mini bullet cluster scenario. 

A potential issue of the mini bullet cluster scenario is the extended distribution of globular clusters (GCs) observed in the DMDGs. As they are massive ($3.8 \times 10^5 - 1.5 \times 10^6 \msun$), dynamical friction causes their orbits to decay \citep[e.g.,][]{Chandrasekhar1943, Nusser2018, Leigh2020}. While core stalling and scattering among GCs can work to suppress the orbital decay \citep{DuttaChowdhury2019, DuttaChowdhury2020}, the impact is limited in the past. Although \cite{Ogiya2022} showed that recursive tidal interactions between NGC~1052 and a progenitor of a DMDG could expand the GC distribution, this would not be the case for the mini bullet cluster scenario. The DM-free gas and forming DMDGs move apart from NGC~1052 as one of the progenitor gas-rich galaxies has a large enough momentum to escape from the host galaxy. Therefore, the GC distribution at the time of formation was more extended than observed at present. The first purpose of this paper is to explore what orbits GCs should be on at the formation epoch to reproduce the observations at present. We address this with a semi-analytic approach. 

The second purpose of this paper is to test if the DMDG can retain GCs on extended orbits under the influence of the tidal force of NGC~1052. In the mini bullet cluster scenario, the GC formation is expected to happen on a short timescale, $\sim 100$\,Myr, immediately following the galaxy collision \citep[][see also e.g. \citealt{Madau2020}]{Lee2021}, and as a consequence, the homogeneous age and metallicity of GCs in DF2 and DF4 may be explained \citep{Fensch2019_df2,vanDokkum2022_gc}. The short timescale of the formation of DMDGs and associated GCs suggests that their birthplace virtually corresponds to the place of the galaxy collision. As one of the progenitor galaxies was a satellite galaxy of NGC~1052, the event should have happened in the potential field of NGC~1052. Thus, GCs belonging to the collision-induced DMDGs are subject to tidal stripping by NGC~1052, depending on the position of the GCs within the DMDG and the location of the galaxy collision. We argue this point based on the analytical model of tidal stripping.

This paper is structured as follows. In \autoref{sec:dev_model}, a semi-analytic model to study the orbital evolution of GCs in an isolated DMDG is developed. Using this model, we study the orbit of GCs at the formation epoch to test the mini bullet cluster scenario from the point of view of the susceptibility to the tidal force of NGC~1052 in \autoref{sec:testing_scenario}. Finally, the results are summarized and discussed in \autoref{sec:summary_discussion}.

%%%%%%%%%%%%%%%%%%%%%%%%%%%%%%%%%%%%%%%%%%%%%%%%%%%
\section{Developing a semi-analytic model}
\label{sec:dev_model}

In this Section, we develop a semi-analytic model to study the orbital evolution of GCs in an isolated DMDG. A fudge parameter in the model is calibrated with results from an \nbody simulation.

%%%%%%%%%%%%%%%%%%%%%%%%%
\subsection{Semi-analytic model}
\label{ssec:semi-ana}

The density structure of the DMDG model is described with two components, an inner core and an outer tail. Each component follows the deprojected \sersic profile,
\begin{equation}
    \rho(r) = \rho\sub{0} \biggl(\frac{r}{R\sub{e}}\biggr)^{-p_n} \exp{\biggl[-b_n \biggl(\frac{r}{R\sub{e}}\biggr)^{1/n} \biggr]}
         \label{eq:deprojected_sersic}
\end{equation}
\citep[e.g.][]{Mellier1987,Prugniel1997}, where $r$ and $\rho\sub{0}$ are the distance from the center of the galaxy and a characteristic density. The effective radius and \sersic index \citep{Sersic1963} are indicated by $R\sub{e}$ and $n$, respectively. We derive the two parameters depending on $n$, $b_n$ and $p_n$, by following the prescriptions by \cite{Ciotti1999} and \cite{LimaNeto1999}. As mentioned above, the total density is given as a sum of the two components,
\begin{equation}
    \rho\sub{tot}(r) = \rho\sub{in}(r) + \rho\sub{out}(r).
        \label{eq:rho_tot}
\end{equation}
Based on the observations of DMDGs, we set $R\sub{e}=2.2$\,kpc and $n=0.6$ for the inner core, $\rho\sub{in}$ \citep{vanDokkum2018_df2}, and the outer tail, $\rho\sub{out}$, is modeled with $R\sub{e}=4.5$\,kpc and $n=1$, i.e. the surface brightness of the outer tail decays exponentially \citep{Montes2020,Keim2021}. While the inner core is dominant at $r < 6$\,kpc, $\rho\sub{out} > \rho\sub{in}$ at larger radii. The dynamical mass of the DMDG model within $r=2.7$ (7.6)\,kpc is $1.4 \times 10^8$ ($3.4 \times 10^8$) $\msun$, consistent with the inference for DF2 \citep{vanDokkum2018_df2,Danieli2019} and the distribution of the material extends up to $r=15$\,kpc. The total dynamical mass of the system is $3.8 \times 10^8 \msun$. The density profile of \autoref{eq:rho_tot} is based on the stellar component of the observed DMDGs while the model includes the little DM component as well and its density profile is assumed to be the same as that of the stellar component. This treatment is justified for DMDGs as the stellar density dominates over the DM density.

In the semi-analytic model, we trace the orbital evolution of GCs under the influence of two forces, the gravity of the global potential of the DMDG and dynamical friction \citep[e.g.][]{Taylor2001}
\footnote{We can neglect core stalling and GC-GC scattering that can prevent GCs from sinking to the center of the DMDG \citep{DuttaChowdhury2019, DuttaChowdhury2020} in our model because of the fundamental difference between their model and ours. On the one hand, the former tracked the orbital evolution of GCs in the future with the forward time integration. The orbital decay due to dynamical friction accumulates GCs in the center of the DMDG with time, making core stalling and GC-GC scattering efficient. On the other hand, as described in the sections below, our semi-analytic model employs the backward time integration and studies the orbital evolution of GCs in the past when those effects were less efficient.}. 
Assuming that the DMDG is spherical, the computation of the former is straightforward,
\begin{equation}
  {\bf a}\sub{global}({\bf r}) = -\frac{GM(r){\bf r}}{r^3},
    \label{eq:a_global}
\end{equation}
where ${\bf r}$ represents the position vector of the GC in the DMDG and $G$ and $M(r)$ are the gravitational constant and the mass enclosed within $r = |{\bf r}|$, respectively. Dynamical friction exerting on the GC is computed with the Chandrasekhar formula \citep{Chandrasekhar1943},
\begin{equation}
  {\bf a}\sub{df}({\bf r}, {\bf v}) =  -4 \pi G^2 M\sub{GC} \ln{\Lambda(r)} \rho(r)f(r,v) \frac{{\bf v}}{v^3},
    \label{eq:a_df}
\end{equation}
where $M\sub{GC}$ and ${\bf v}$ are the mass and the velocity vector of the GC, respectively. Assuming the Maxwell-Boltzmann velocity distribution of materials in the DMDG, the fraction of materials moving with a velocity less than $v = |{\bf v}|$, that contribute to the process of dynamical friction, is given by
\begin{equation}
  f(r,v) = {\rm erf}\biggl [ \frac{v}{\sqrt{2}\sigma(r)} \biggr ]  - \sqrt{\frac{2}{\pi}}\frac{v}{\sigma(r)}\exp{\biggl [- \frac{v^2}{2\sigma(r)^2} \biggr ]}, 
    \label{eq:f_v}
\end{equation}
where $\sigma(r)$ is the velocity dispersion at $r$. The radial profiles of $M(r)$ and $\sigma(r)$ are numerically derived based on \autoref{eq:rho_tot}. We employ the Coulomb logarithm depending on $r$ \citep{Hashimoto2003},
\begin{equation}
  \ln{\Lambda(r)} = \ln(r/b\sub{min}),
       \label{eq:Coulomb_log}
\end{equation}
where $b\sub{min}$ is a parameter and we calibrate it using an \nbody simulation in \autoref{ssec:calibration}. The GC orbit is integrated with a second-order accuracy about the timestep, $\Delta t$. Throughout the paper, we fix the timestep of the semi-analytic model as $\Delta t = 1$\,Myr. Experiments varying $\Delta t$ confirm that the results are converged. Note that $M\sub{GC}$ is supposed to be constant in the model.

%%%%%%%%%%%%%%%%%%%%%%%%%
\subsection{N-body simulation}
\label{ssec:nbody}

To calibrate the parameter in the semi-analytic model, $b\sub{min}$, we perform an \nbody simulation of an isolated DMDG that contains ten GCs. Its density structure follows \autoref{eq:rho_tot} and we use the acceptance-rejection sampling method \citep{Press2002} to draw the initial position and velocity vectors of \nbody particles. The distance from the center of the DMDG to a particle, $r$, is sampled based on \autoref{eq:rho_tot}. We also randomly draw a unit vector to specify the three-dimensional position of the particle. The phase-space distribution function is computed using the Eddington formula \citep{Eddington1916} to sample the norm of the velocity vector of a particle, $v$. As the phase-space distribution function is assumed to depend only on energy, we specify the three-dimensional velocity vector of the particle with $v$ and another randomly drawn unit vector. 

Ten particles are selected with the following procedure and treated as GCs in the simulation. We consider a shell with a radius of $r\sub{GC}$ and find the ten closest particles to the shell. While their position and velocity stay as drawn by the acceptance-rejection sampling method, we increase their mass to $10^6 \msun$. Although this selection scheme is somewhat artificial, GC particles spread out to a projected spatial distribution consistent with those in observed galaxies in 100\,Myr \citep{Ogiya2022}. In this study, we set $d\sub{GC}=8$\,kpc as the size of the GC distribution and the velocity dispersion of GCs are reasonably consistent with observations after the dynamical evolution of 8\,Gyr.

We perform the \nbody simulation using a tree code \citep{Barnes1986}, developed for Graphics Processing Units clusters \citep{Ogiya2013}. The cell opening criteria of \cite{Springel2005_gadget2} with the parameter controlling the force accuracy of $\alpha=0.01$ is employed. The DMDG is modeled with $256^3$ particles and each particle has a mass of $\sim 22 \msun$, while the mass of ten GC particles is $10^6 \msun$. The gravitational potential field is softened with a Plummer \citep{Plummer1911} force softening parameter of 10\,pc. The particle orbit is integrated with the second-order Leapfrog scheme, and the timestep is updated with the prescription of \cite{Power2003} and is equal for all particles. We confirm that the simulation results are numerically converged with simulations varying the number of particles or the softening parameter.

%%%%%%%%%%%%%%%%%%%%%%%%%
\subsection{Calibration of $b\sub{min}$}
\label{ssec:calibration}

%%%%%%%%%%
\begin{figure}
    \begin{center}
        \includegraphics[width=0.4\textwidth]{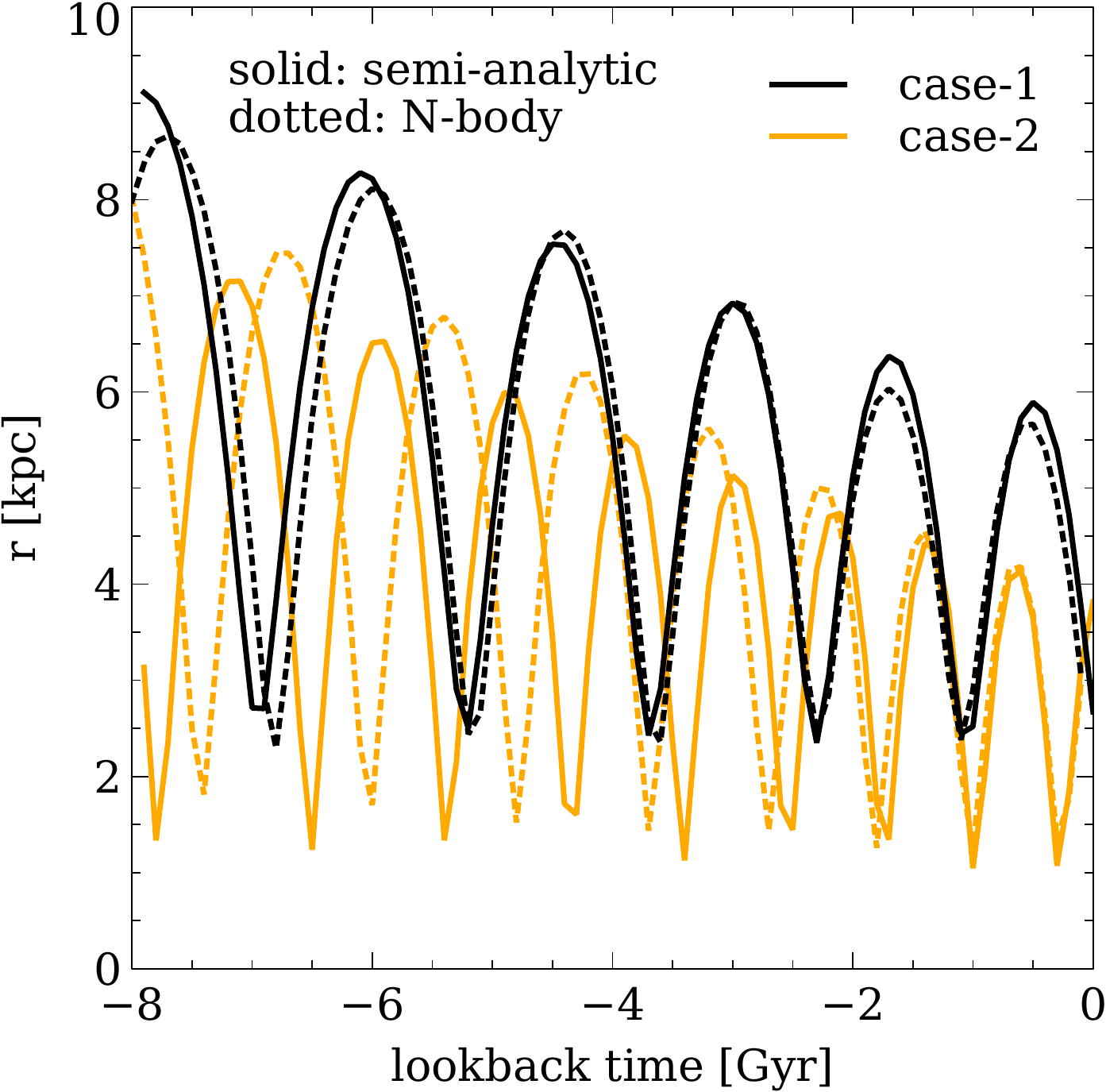}
    \end{center}
    \caption{
        Orbital evolution of two GCs in the DMDG model (black and orange). Solid and dotted lines represent the prediction by the semi-analytic model and the \nbody simulation result, respectively. While the simulation employs the forward time integration, the semi-analytic model adopts the last snapshot of the GCs in the simulation as the {\it final condition} and traces back their orbital evolution with the backward time integration. The semi-analytic model reasonably reproduces the simulation with $b\sub{min}=1$\,pc. 
    \label{fig:gc_orbit}}
\end{figure}
%%%%%%%%%%

The semi-analytic modeling aims to study the orbits that the GCs in an isolated DMDG should have been on at the time of their formation ($t=-8$\,Gyr) to reproduce the observations at present ($t=0$). Given the position and velocity of ten GCs in the last snapshot from the \nbody simulation, {\it final condition}, we trace back their orbital evolution in the isolated DMDG to $t=-8$\,Gyr. Note that the \nbody simulation is performed from $t=-8$\,Gyr to 0. 

In \autoref{fig:gc_orbit}, we compare the prediction by the semi-analytic model (solid) to the results from the \nbody simulation (dotted). The orbital evolution of two GCs is shown (black and orange). After some experiments varying the parameter, $b\sub{min}$, we find $b\sub{min}=1$\,pc reasonably reproduces the simulation results. Typically, the difference between the semi-analytic model and the \nbody simulation in the orbital energy and angular momentum of GCs is less than 10 percent at $t=-8$\,Gyr. We also find that a constant Coulomb logarithm of $\ln{\Lambda}=8$ yields the same level of precision in the orbital energy and angular momentum of GCs at $t=-8$\,Gyr and the results shown in \autoref{sec:testing_scenario} are insensitive to the choice of the Coulomb logarithm.

%%%%%%%%%%%%%%%%%%%%%%%%%%%%%%%%%%%%%%%%%%%%%%%%%%%
\section{Testing the mini bullet cluster scenario}
\label{sec:testing_scenario}

This section aims to test the mini bullet cluster scenario for forming the trail of diffuse galaxies, including two DMDGs, by assessing the susceptibility of GCs in the DMDG to the tidal force of the host galaxy. We first derive the GC orbit at the formation epoch of the DMDG using the semi-analytic model in \autoref{ssec:gc_orbit}. Then, \autoref{ssec:tidal_susceptibility} is devoted to arguing how the GCs are susceptible to the tidal force.

%%%%%%%%%%%%%%%%%%%%%%%%%
\subsection{Orbits of GCs at the formation epoch}
\label{ssec:gc_orbit}

We use the semi-analytic model to study what orbits the GCs should have been on at the formation epoch to reproduce the observed position and velocity at present. While three of six phase-space coordinates ($x$, $y$ and $v\sub{z}$)\footnote{In the observations, the projection plane defines the coordinates of $x$ and $y$, while the direction of the line-of-sight defines the $z$-axis.} as well as the mass, $M\sub{GC}$, of ten GCs in DF2 have been observationally obtained \citep{DuttaChowdhury2019}, the remaining three phase-space coordinates ($z$, $v\sub{x}$ and $v\sub{y}$) are unconstrained. Thus we stochastically sample these quantities using the scheme outlined in \cite{DuttaChowdhury2019}. The projected spatial distribution of GCs at present is described by the \sersic profile of $n=1$ and $R\sub{e}=3.1$\,kpc, assuming that the GC distribution is spherically symmetric. We draw $z$ of each GC based on \autoref{eq:deprojected_sersic}, while $x$ and $y$ are given as observed. The distribution of $v\sub{z}$ of GCs in DF2 is described by the Maxwell-Boltzmann distribution of $\sigma=7.8$\,km/s. Assuming that the velocity dispersion of the GC population is isotropic, we sample $v\sub{x}$ and $v\sub{y}$ from the same distribution, while $v\sub{z}$ of each GC is set as observed. For each GC, $10^6$ random realizations are studied, and we consider $10^7$ cases in total.

Using the semi-analytic model, we measure the maximum distance from the center of the DMDG to GCs, $r\sub{max}$. This is a crucial quantity to argue the susceptibility of GCs to the tidal force of the host galaxy, as materials in the outskirt of the satellite galaxies are more easily stripped compared with those in the satellite center (see \autoref{ssec:tidal_susceptibility}). As dynamical friction decays orbits of GCs, $r\sub{max}$ depends on the measuring time. According to hydrodynamic simulations of the mini bullet cluster scenario \citep{Shin2020, Lee2021}, DMDGs are formed in $\sim 100-1000$\,Myr, depending on the collision orbit. Thus, we select two time windows of $t=[-8, -7.5]$ and $[-7.5, -7]$\,Gyr to measure $r\sub{max}$ in the analysis. 

%%%%%%%%%%
\begin{figure}
    \begin{center}
        \includegraphics[width=0.4\textwidth]{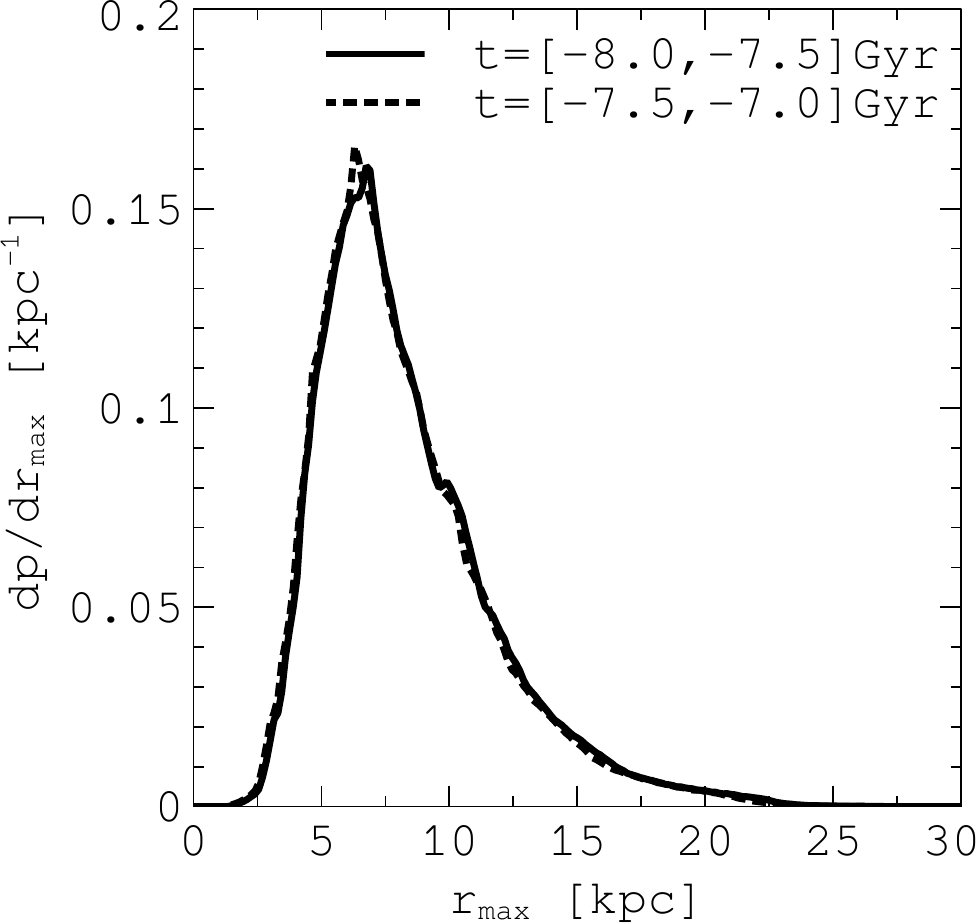}
    \end{center}
    \caption{
      Probability distribution of the maximum-$r$, $r\sub{max}$, at $t=[-8, -7.5]$\,Gyr (solid) and at $t=[-7.5,-7]$\,Gyr (dotted). GCs should have been on orbits of $r\sub{max} = 5-10$\,kpc at the formation epoch.
    \label{fig:prmax}}
\end{figure}
%%%%%%%%%%

In \autoref{fig:prmax}, we show the probability distribution of $r\sub{max}$ measured at $t=[-8, -7.5]$\,Gyr (solid) and at $t=[-7.5,-7]$\,Gyr (dotted). The $r\sub{max}$-distribution is unchanged in the first Gyr of the evolution. We find that GCs are likely to have $r\sub{max}=5-10$\,kpc at the formation epoch of the DMDG. While there is a long tail on the larger $r\sub{max}$ side, the distribution sharply decays on the smaller $r\sub{max}$ side.

%%%%%%%%%%%%%%%%%%%%%%%%%
\subsection{Susceptibility of GCs to the tidal force}
\label{ssec:tidal_susceptibility}

In this subsection, we assess how GCs in the DMDG are susceptible to tidal force by combining the results from the semi-analytic modeling and the analytic model of tidal stripping. The mean density, ${\bar \rho}$, is a useful indicator to argue the susceptibility of a satellite galaxy to the tidal force. When the mean density of the host, ${\bar \rho}\sub{host}$, exceeds that of the satellite, ${\bar \rho}\sub{sat}$, the material contained in the satellite will be stripped by the tidal force of the host galaxy. The mean density is a function of the distance from the center of the system, $d$, ${\bar \rho}(d) \equiv 3M(d)/ 4 \pi d^3$, where $M(d)$ is the enclosed mass within $d$. In the analysis, the mass profile of the DMDG is based on \autoref{eq:rho_tot}. We suppose that the density structure of NGC~1052 is described by the Navarro-Frenk-White model \citep[NFW;][]{Navarro1997}\footnote{The tidal interaction can, in fact, compress the DMDG and GCs will not be stripped when the density profile of NGC~1052 is shallower than ${\bar \rho\sub{host}} \propto R^{-1}$ \citep{Dekel2003}. As the NFW profile is steeper than the above-mentioned critical slope at all radii, the tidal interaction works as a stripping process. This is the same when the central stellar component, which is well described by the Hernquist profile \citep{Hernquist1990}, is included.} and its structural parameters (virial mass of $M\sub{200}=5.1 \times 10^{12} \msun$ and concentration of $c=5.3$) are derived by empirical relations obtained from cosmological \nbody simulations \citep{Correa2015,Ludlow2016}, assuming redshift of $z=1$ (corresponding lookback time is 8\,Gyr) and the current virial mass of the galaxy, $1.1 \times 10^{13}\msun$ \citep{Forbes2017,Behroozi2019}. Employing the cosmological parameter set of \cite{Planck2016}, the virial radius, in which the mean density is 200 times the critical density of the universe at $z=1$, of NGC~1052 is $R\sub{200}=246$\,kpc.

%%%%%%%%%%
\begin{figure}
    \begin{center}
        \includegraphics[width=0.4\textwidth]{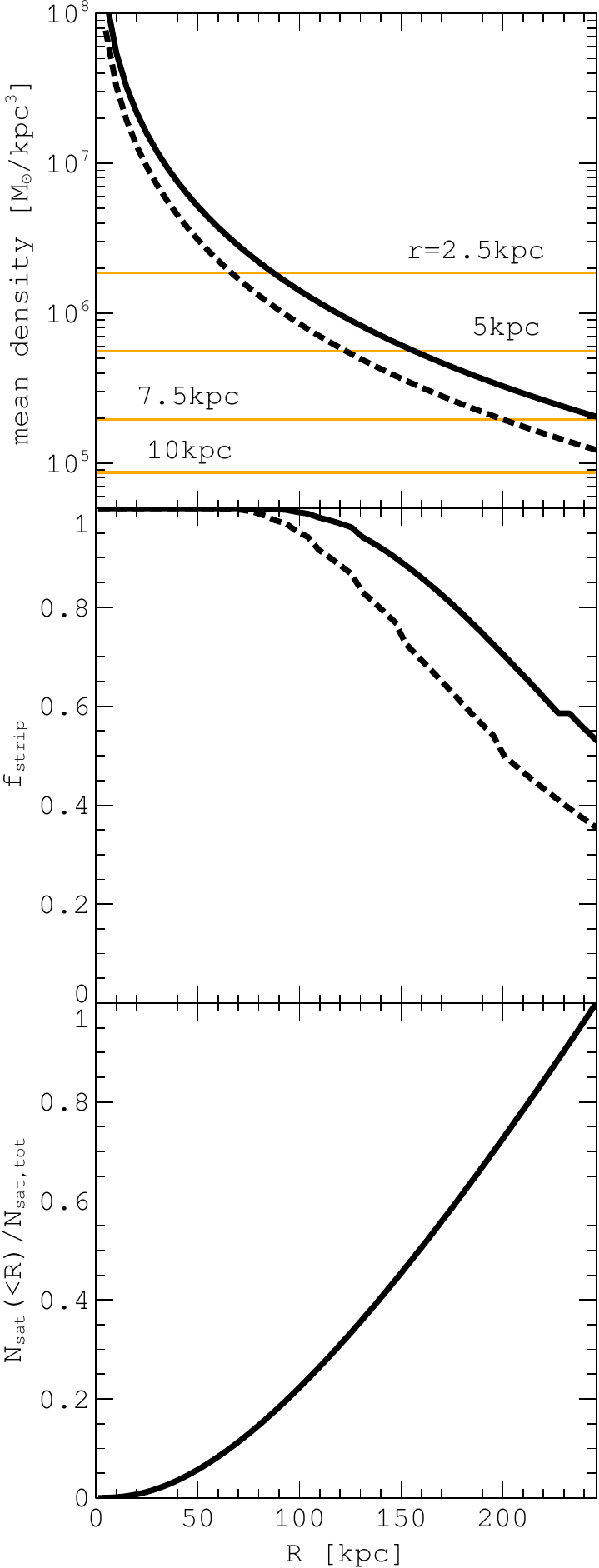}
    \end{center}
    \caption{
        ({\it Top}) Comparison of mean densities. Orange lines show the mean density of the DMDG measured at the indicated $r$. The mean density of NGC~1052 is given as a function of the distance from its center, $R$, and multiplied by a factor of 2.5 (solid black) or 1.5 (dotted black). 
        ({\it Middle}) Fraction of GCs stripped from the DMDG when it is at $R$. 
        ({\it Bottom}) Cumulative number fraction of satellite galaxies contained within $R$. The NFW density profile with parameters explained in the main text and the power-law index of $\beta=1$ are used to compute it. Half of the satellite galaxies are located at $R < 150$\,kpc where the tidal force reduces the number of GCs by a factor of 5-10. 
    \label{fig:tidal_stripping}}
\end{figure}
%%%%%%%%%%

The orange horizontal lines in the top panel of \autoref{fig:tidal_stripping} represent the mean density of the DMDG, ${\bar \rho}\sub{sat}$, at $d = r$ as indicated. We compute the mean density of NGC~1052, ${\bar \rho}\sub{host}$, as a function of the distance from its center, $d = R$. The solid and dotted black curves show $2.5{\bar \rho}\sub{host}$ and $1.5{\bar \rho}\sub{host}$, respectively. The pre-factor comes from the analytical model of tidal radius, $r\sub{t}$, 
\begin{eqnarray}
    {\bar \rho\sub{sat}}(r\sub{t}) &=& [\alpha - d\ln{M\sub{host}}/d\ln{R}|\sub{R}] {\bar \rho\sub{host}}(R) \nonumber \\ 
                                   &\equiv& \alpha' {\bar \rho\sub{host}}(R),
        \label{eq:tidal_radius}
\end{eqnarray}
where $M\sub{host}$ is the mass profile of the host galaxy. Models of \cite{King1962} and \cite{Tormen1998} indicate $\alpha=3$ and 2, respectively.\footnote{A comprehensive review is found in \cite{vandenBosch2018a}.} Assuming the NFW model, the logarithmic slope of the mass profile is $\sim 0.5$ at the virial radius of the host galaxy. The pre-factor of $\alpha'=1.5-2.5$ is also consistent with \cite{Drakos2022}. 

We derive the fraction of GCs to be stripped, $f\sub{strip}$, by combining the analytic model of tidal radius and the $r\sub{max}$-distribution (\autoref{ssec:gc_orbit}). The tidal massloss of satellite galaxies in a single orbit is estimated with the instant tidal radius measured at the closest approach to the host in the orbit \citep[e.g.][]{Penarrubia2010,vandenBosch2018a}. Assuming that the closest approach of the DMDG to the host galaxy is $R$, we compute a critical distance from the center of the DMDG, $r\sub{crit}$, satisfying the condition of \autoref{eq:tidal_radius}, i.e. $r\sub{crit} = r\sub{t}$, and weight $r\sub{max} \geq r\sub{crit}$ with the $r\sub{max}$-distribution to derive $f\sub{strip}$. The middle panel of \autoref{fig:tidal_stripping} shows $f\sub{strip}$ as a function of the location of the DMDG in NGC~1052, $R$, and indicates that if DMDGs were located at $R < 150$\,kpc, more than 80 percent of GCs will be lost from the DMDG. While the fraction gets lower at larger $R$, more than half of GCs are expected to be stripped at the virial radius of NGC~1052 at $z=1$ ($R\sub{200} = 246$\,kpc).  

The location of the DMDG formation within the host galaxy is a critical factor in determining the fate of GCs orbiting within the DMDG. Since one of the progenitors of the collision-induced DMDGs is a satellite galaxy bound to NGC~1052, the location of the galaxy collision can be estimated from the distribution of satellite galaxies. Cosmological \nbody simulations have studied the spatial distribution of DM substructures \citep[][and references therein]{Ghigna2000,Nagai2005,Gao2012}. \cite{Han2016} found that the number density profile of them can be modeled as a product of the density profile of the host, $\rho\sub{host}(R)$ (e.g., NFW model) and a power-law, $R^{\beta}$ with $\beta \approx 1$. Assuming that DM substructures hosting satellite galaxies are distributed spherically in the host, the number of satellites located at $R$ is proportional to $dN\sub{sat}/dR \propto \rho\sub{host}(R) R^{\beta+2}$. In the bottom panel of \autoref{fig:tidal_stripping}, we present the number fraction of satellite galaxies contained within $R$ and find that half of the satellites are located in the central 150\,kpc where 80-90 percent of GCs will be stripped from the DMDG.

How many GCs are stripped by the tidal force? To address this question, we assume that the distributions of $dp/dr\sub{max}$ and $dN\sub{sat}/dR$ are uncorrelated with each other and construct a two-parameter distribution, $d^2p/(dr\sub{max}dR)$. Weighing the pairs of $(r\sub{max}, R)$ with this distribution, we find that 83 (70) percent of GCs will be tidally stripped from the DMDG when assuming $\alpha'=2.5$ (1.5). As ten GCs are observed in DF2, it turns out that 33-59 GCs should be originally formed in the collision-induced DMDGs. While 42 star clusters were formed in the hydrodynamic simulation of the mini bullet cluster scenario \citep{Lee2021}, half are less massive than GCs observed in DF2. The observations might be explained if less massive star clusters are formed at larger radii where they will be selectively stripped, although the simulation did not show such distribution. Therefore, the number of massive GCs formed in the mini bullet cluster scenario is a potential issue.

%%%%%%%%%%%%%%%%%%%%%%%%%%%%%%%%%%%%%%%%%%%%%%%%%%%
\section{Summary and discussion} 
\label{sec:summary_discussion}

Recently, \cite{vanDokkum2022_trail} suggested that the trail of galaxies with low surface brightness, including two DMDGs (DF2 and DF4), in the vicinity of NGC~1052 might have been formed through a high-speed collision between two gas-rich dwarf galaxies at $z \sim 1$ (8\,Gyr ago). A burst of star formation activity is induced due to the strong compression of the galactic gas. As the DM and pre-existing stars in the progenitor galaxies pass through the gas, the stars formed in the compressed gas can create DMDGs \citep{Silk2019, Trujillo-Gomez2021}. The DM-free gas subsequently fragments to form a trail of diffuse galaxies.  

A challenge for the mini bullet cluster scenario is the extended distribution of GCs (a few kpc in projection) in the observed DMDGs. The orbit of GCs has been shrunk due to dynamical friction. Thus their distribution at the time of formation was more extended than at present. Using a semi-analytic model, we find that the observed position and velocity of GCs can be reproduced if they were at $r=5-10$\,kpc at the formation epoch. As the mini bullet cluster model predicts that the DMDGs and associated GCs are formed immediately after the galaxy collision near NGC~1052, GCs are subject to tidal stripping by the host galaxy. Combining the GC distribution at the formation epoch with the analytic models of tidal radius and the distribution of satellites, we find that 70-83 percent of GCs should have been stripped from the DMDG. More than 33-59 GCs need to be originally formed to explain the observed number of GCs in DF2 (ten). While $\sim 40$ star clusters could be formed in the scenario, half of them are less massive than GCs in DF2. The simulation did not find a tendency for less massive star clusters to be distributed at larger radii where they can be more easily stripped from the DMDG. Therefore, the number of massive GCs is a potential issue for the mini bullet cluster scenario.

A caveat on our argument is that the semi-analytic model considers the orbital evolution of GCs in an isolated DMDG, while DMDGs formed in the mini bullet cluster scenario are expected to be under the influence of the tidal force of the host galaxy. One may suppose that GCs were initially on compact orbits, preventing tidal stripping, and the injection of kinetic energy through the process of tidal shock \citep[e.g.][]{Spitzer1958, Gnedin1999, Banik2021_tidal} can expand the GC orbits to the level mentioned above. While \cite{Ogiya2022} showed that multiple pericentric passages are needed to reproduce the extended distribution of GCs, the DMDGs formed in the mini bullet cluster scenario considered in \cite{vanDokkum2022_trail} interact with the host only once at the time of formation. Then they move apart from the host galaxy. Thus, the injection of kinetic energy by tidal shock does not help to maintain the extended GC distribution in the mini bullet cluster scenario.

Hydrodynamic simulations of the mini bullet cluster scenario formed DMDGs, together with tens of star clusters \citep{Lee2021}. However, several challenges remain to explain observations of DMDGs with the scenario. First, the host galaxy was absent in those simulations, while it plays a role in modifying the properties of satellite galaxies. Our analysis shows that a significant fraction of GCs can be lost from the DMDG interacting with the host galaxy. Second, DMDGs formed in the simulations are too compact ($\la 0.1$\,kpc) compared to the observed ones \citep{Shin2020, Lee2021}. They need expansion processes to reproduce observations. In this respect, the interactions with the host galaxy may be essential, and satellite galaxies can be more efficiently puffed up on more radial orbits. However, in such cases, the satellites approach the host center, and the ram pressure of the host gas removes the gas from the satellite before the galaxy collision. Finally, while investigating if the DMDG models can transform into ultra diffuse galaxies, like DF2 and DF4, is interesting, the simulated time in the previous studies ($<1$\,Gyr) is short for discussing the evolution after their formation. The outcome would depend on the interaction configuration among three galaxies (two gas-rich dwarfs and the host galaxy). An extensive parameter survey of high-resolution numerical simulations with a long enough simulation time is a promising way to achieve a firm conclusion.

%% Also note that the akcnowlodgment environment does not support long amounts of text. If you have a lot of people and institutions to acknowledge, do not use this command. Instead, create a new \section{Acknowledgments}.
\begin{acknowledgments}
We thank the anonymous referee for providing us with insightful comments that improved the article. GO and XK acknowledge the Fundamental Research Fund for Chinese Central Universities (No. NZ2020021 and No. 226-2022-00216), NSFC (No. 11825303, 11861131006), the science research grants from the China Manned Space project (No. CMS-CSST-2021-A03, CMS-CSST-2021-B01), and the cosmology simulation database in the National Basic Science Data Center (NBSDC) and its funds the NBSDC-DB-10. FvdB is supported by the National Aeronautics and Space Administration through Grant No. 19-ATP19-0059 issued as part of the Astrophysics Theory Program. AB acknowledges support from the Excellence Cluster ORIGINS which is funded by the Deutsche Forschungsgemeinschaft (DFG, German Research Foundation) under Germany's Excellence Strategy-EXC-2094-390783311.

\end{acknowledgments}

\bibliography{dmdg_gc}{}
\bibliographystyle{aasjournal}



\end{document}